\documentclass[sigconf]{acmart}
\AtBeginDocument{%
  }

\setcopyright{acmlicensed}
\copyrightyear{2025}
\acmYear{2025}
\setcopyright{cc}
\setcctype{by}
\acmConference[CHI '25]{CHI Conference on Human Factors in Computing Systems}{April 26-May 1, 2025}{Yokohama, Japan}
\acmBooktitle{CHI Conference on Human Factors in Computing Systems (CHI '25), April 26-May 1, 2025, Yokohama, Japan}\acmDOI{10.1145/3706598.3714074}
\acmISBN{979-8-4007-1394-1/25/04}

\usepackage{subfigure}
\usepackage{makecell}
\usepackage{enumitem}
\begin{document}

\title[Raising Awareness of Location Information Vulnerabilities in Social Media Photos using LLMs]{Raising Awareness of Location Information Vulnerabilities in Social Media Photos using LLMs}
\author{Ying Ma}
\orcid{0000-0001-5413-0132}
\affiliation{%
  \department{School of Computing and \\Information Systems}
  \institution{University of Melbourne}
  \city{Melbourne}
  \country{Australia}}
  \email{ying.ma1@student.unimelb.edu.au}

\author{Shiquan Zhang}
\orcid{0000-0003-3747-0842}
\affiliation{%
  \department{School of Computing and \\Information Systems}
  \institution{University of Melbourne}
  \city{Melbourne}
  \country{Australia}}
\email{shiquan.zhang@student.unimelb.edu.au}

\author{Dongju Yang}
\orcid{0000-0003-1657-7130}
\affiliation{%
 \department{School of Computing and \\Information Systems}
 \institution{University of Melbourne}
 \city{Melbourne}
  \country{Australia}
    }
  \email{dongjuy@student.unimelb.edu.au}

\author{Zhanna Sarsenbayeva}
\orcid{0000-0002-1247-6036}
\affiliation{%
 \department{School of Computer Science}
 \institution{University of Sydney}
 \city{Sydney}
  \country{Australia}
    }
\email{zhanna.sarsenbayeva@sydney.edu.au}

\author{Jarrod Knibbe}
\orcid{0000-0002-8844-8576}
\affiliation{%
 \department{School of Electrical Engineering and Computer Science}
 \institution{University of Queensland}
 \city{St Lucia}
  \country{Australia}
    }
      \email{j.knibbe@uq.edu.au}

\author{Jorge Goncalves}
\orcid{0000-0002-0117-0322}
\affiliation{%
 \department{School of Computing and \\Information Systems}
 \institution{University of Melbourne}
 \city{Melbourne}
  \country{Australia}
  }
\email{jorge.goncalves@unimelb.edu.au}

\renewcommand{\shortauthors}{Ma et al.}
\begin{abstract}


Location privacy leaks can lead to unauthorised tracking, identity theft, and targeted attacks, compromising personal security and privacy. This study explores \textcolor{black}{LLM-powered location privacy leaks} associated with photo sharing on social media, focusing on user awareness, attitudes, and opinions. We developed and introduced \textcolor{black}{an LLM-powered location privacy intervention} app to 19 participants, who used it over a two-week period. The app prompted users to reflect on potential privacy leaks \textcolor{black}{that a widely available LLM could easily detect}, such as visual landmarks~\& cues that could reveal their location, and provided ways to conceal this information. Through in-depth interviews, we found that our intervention effectively increased users' awareness of location privacy \textcolor{black}{and the risks posed by LLMs}. It also encouraged users to consider the importance of maintaining control over their privacy data and sparked discussions about the future of location privacy-preserving technologies. Based on these insights, we offer design implications to support the development of future user-centred, location privacy-preserving technologies for social media photos.

\end{abstract}




\maketitle

\section{Introduction}

Social Network Sites (SNSs) have become integral to how people connect, communicate, and share their lives with others. These platforms provide users with the ability to manage their online impressions, maintain social relationships, and even gain broader attention through the content they share~\cite{oeldorf2010online}. Among the various activities on these platforms, photo sharing has surged in popularity, with billions of images exchanged daily. The abundance of SNSs has made it easier than ever to share visual content quickly and effortlessly~\cite{cho2019effects}. However, despite this convenience, many users remain largely unconcerned or unaware of the privacy risks associated with sharing their photos online. 
\textcolor{black}{Furthermore, information overload contributes to privacy fatigue, which can leave users indifferent to these risks~\cite{shao2022you}.}
This growing desensitisation to privacy issues on social media has led to a significant increase in unintentional location data leaks through shared images, resulting in notable privacy breaches~\cite{dehart2020social}.

Importantly, among the different types of privacy leaks, those related to location privacy are particularly concerning. Whether location data is shared intentionally through tagging or unintentionally through metadata, the consequences can be severe, such as instances of doxing and stalking ~\cite{cohen2022internet}. Furthermore, the rise of sophisticated tools such as large language models (LLMs) and Google Lens, \textcolor{black}{which can recognise landmarks, street signs, and other contextual elements in images,} has made it easier to detect photo locations without users' awareness, significantly threatening location privacy on social media. \textcolor{black}{Our study specifically focuses on the risks associated with LLMs now being able to quickly and accurately extract location information from just visual content in photos~\cite{yang2024geolocator}, rather than explicit location data such as geotags or location tags deliberately added during sharing. This distinction is critical to understanding how users perceive their location privacy within the context of unintentional information disclosure.}

Most existing approaches to preserving location privacy focus on technical solutions, including obfuscation-based, cryptography-based, and cooperation and caching-based mechanisms~\cite{jiang2021location}. However, there is a significant gap in addressing this issue from a Human-Computer Interaction (HCI) perspective~\cite{liu2022privacy} in terms of raising users' awareness of the privacy risks stemming from everyday photo sharing, \textcolor{black}{particularly in a time where increasingly more powerful LLMs can be used to extract sensitive location information from photos}. Awareness-focused interventions are crucial, as technical solutions alone may not effectively change the behaviours that lead to unintentional location data exposure on social media. By enhancing users' understanding of potential location privacy leaks, they may be inclined to adopt more protective practices~\cite{thakkar2022would}.

To explore this further, we designed and implemented a location privacy-preserving intervention app. The app aims to heighten users' awareness of potential privacy leaks in the photos they intend to share online by \textcolor{black}{showing them how a widely available LLM can easily} detect the location where the photos were taken. It also provides users with editing features---such as blurring and cropping---that allow them to obscure or remove location-revealing information, as well as gives location privacy warnings, empowering users to take control of their privacy. We conducted a 2-week user study that used our app, and sought to understand how users perceive this tool and how it influenced users' attitudes towards location privacy. 
\textcolor{black}{ Our work addresses the following research questions:}
\begin{itemize}
    \item
    \textcolor{black}{\textbf{RQ1:} What is the impact of our intervention app on participants' awareness of the risks associated with LLM-powered location information leaks in photos?}
    \item
    \textcolor{black}{\textbf{RQ2:} How does this intervention influence participants' intentions to manage their location privacy practices?}
    \item\textcolor{black}{\textbf{RQ3:} What features could be introduced to future location privacy-preserving tools and approaches to mitigate concerns regarding LLM-powered leaks?}
\end{itemize}

Our findings suggest that when users are made aware of the potential for location privacy leaks in their photos, they \textcolor{black}{express an intention to adjust their photo-taking and sharing practices.} Our participants also found the app's editing features, in particular, to be effective in helping obscure location-revealing information, thereby reducing their risk of unintended location exposure. Furthermore, participants generally perceived the app as a valuable tool for privacy protection, appreciating how it enabled them to have a greater sense of control over their personal information. In addition, participants offered suggestions for future location privacy-preserving technologies to help alleviate users' concerns and encourage better engagement.
Our study makes the following contributions:
\begin{itemize}
    \item We provide an in-depth investigation of location privacy risks associated with photo sharing on social media with the advent of LLM technologies. We highlight how information within photos can easily lead to location privacy leaks, and how helping users easily identify these issues can raise their awareness.

    \item We identify scenarios where users would like more control over their location privacy, and reveal how users \textcolor{black}{consider ways to} adjust their behaviour in response to these varying risks.
    
    \item Our work provides insights into how location privacy-preserving tools can better support users in protecting their location information. We offer design recommendations for how tools can be developed to enhance user trust, improve usability, and encourage proactive engagement in managing their online location privacy.
    

\end{itemize}


\section{Related Work}

In this section, we review existing literature on privacy risks \textcolor{black}{and awareness} in social media platforms, with a particular attention to the challenges posed by sharing photos. We then examine research that explores location privacy concerns related to photo sharing. Finally, we discuss \textcolor{black}{several attack methods that encroach on users' location privacy}, and summarise mechanisms or interventions that have been proposed with the aim of protecting users' location privacy.

\subsection{\textcolor{black}{Privacy Risks and Awareness in Social Media Platforms}}

Privacy leaks associated with social media platforms have emerged as critical concerns in today's digital landscape. Many individuals unknowingly disclose sensitive information through their posts, photos, and interactions, \textcolor{black}{often due to a lack of awareness about the potential privacy implications~\cite{nyoni2018privacy,henne2013awareness}.} 
This phenomenon is exacerbated by the design of social media applications, which frequently encourage users to post and share for social validation or engagement, leading to inadvertent privacy breaches~\cite{jain2021online}.
\textcolor{black}{Furthermore, users tend to underestimate the size of their audience, leading to oversharing and unexpected privacy leaks \cite{bernstein2013quantifying}. Even if a user is privacy-conscious and accurately estimates the reach of their posts, their network could still disclose information about them~\cite{bagrow2019information}.
This lack of user awareness highlights the need for more effective strategies to educate users about privacy risks and promote protective behaviours.}

\textcolor{black}{To address these challenges, efforts have been made to raise awareness regarding privacy on SNSs. For instance, 
systems should increase users' awareness and minimise the effort required for reconfiguration by enabling users to effortlessly and regularly update their privacy settings as their social relationships evolve~\cite{li2018sns}.
Another study conducted by~\citet{kroll2021digital} demonstrates how
Facebook utilises strategies like simplification and reminders to increase users' awareness of privacy settings. However, their study also indicates that nudges designed to increase privacy awareness do not always produce clear outcomes, as they may be quickly forgotten or not taken seriously by users at the outset.}
\textcolor{black}{\citet{assal2015s} identify the need to address issues such as lack of feedback, inadequate abstraction properties, and unmotivated user behaviour, which contribute to low adoption rates of privacy-preserving features on SNSs.}

\textcolor{black}{Moreover, the complexity of privacy settings and the challenge of managing non-standardised configurations across different contexts impose significant cognitive demands on users~\cite{seberger2021empowering,ma2024understanding}. 
This often leads to ``privacy fatigue'', a state of exhaustion that reduces protective intention, even when users are aware of privacy risks~\cite{choi2018role}. This fatigue significantly impacts users' willingness to engage in privacy-protective measures, highlighting the need for more intuitive and uniform privacy controls~\cite{jafari2023navigating}. 
As users become increasingly fatigued, they may become less likely to take proactive steps to protect their personal information}. 

\subsection{Location Privacy Concerns when Sharing Photos Online}
Photos can capture significant life moments \textcolor{black}{and often contain detailed information about the individuals depicted, their activities, objects, and the surrounding environment ~\cite{kairam2016snap,mcgookin2019reveal}.
  However, this rich contextual information significantly increases the risk of privacy breaches, as users may inadvertently disclose private details through shared images, raising substantial concerns about visual privacy~\cite{dehart2020social}. 
Previous work has explored photo privacy using a human-centred taxonomy, identifying 28 categories of sensitive content in photos, including appearance, facial expressions, medical conditions, and personally identifiable information~\cite{li2020towards}. While users might be unaware of the potential consequences of sharing photos online~\cite{henne2014study}}, they often prioritise the desire for social connectivity and self-expression over privacy considerations, and this tendency drives individuals to share personal information, images, and messages without fully assessing the associated privacy implications~\cite {darwish2019photos,malik2016uses}. 


Regarding all types the risks of photos shared online, location data is particularly important as it affects users' safety and security. It can introduce bias ~\cite{ma2025exploring} and potentially expose individuals' real-time address or daily routines, making them susceptible to tracking, theft, or physical harm~\cite{ma2023hello}. 
\textcolor{black}{With the increasing incorporation of location-based features in photos and SNSs, images often include GPS coordinates in metadata that can be used to pinpoint specific locations.}
\textcolor{black}{A common scenario occurs when a user tags their location in a photo post and includes another individual, either by tagging them or featuring their face in the photo, inadvertently disclosing the location of both parties. 
To protect themselves from such issues,} some users resort to drastic measures like deactivating or deleting their social media accounts. However, this does not prevent the creation of shadow profiles where information about them is shared widely by their networks~\cite{garcia2018collective}.
These potential risks highlight the importance of understanding the implications of photo sharing in digital spaces, particularly with regards to location privacy.

\textcolor{black}{Previous studies have investigated the influence of location on privacy decisions in photo sharing.}
An early study \textcolor{black}{conducted} by~\citet{ahern2007over} analysed 36,000 photos on the photo-sharing mobile app ZoneTag and found that the decision to share a photo privately or publicly was often influenced by the location where the photo was taken, \textcolor{black}{as some locations were perceived as more private than others.}
Participants expressed heightened concerns about revealing granular location details, indicating the sensitive nature of location information in photo sharing. 
\textcolor{black}{Similarly, ~\citet{wu2011temporal} found that users were less likely to tag location on photos taken in private locations compared to public ones.
Another challenge raised by ~\citet{shu2018cardea} is the dynamic and temporary nature of privacy preferences. For example, an individual might typically have no privacy concerns about a location but suddenly wish to remain anonymous in a photo taken there.}

To summarise, the issue of location leakage through image sharing is particularly noteworthy and introduces safety risks~\cite{hoyle2015sensitive}.  \textcolor{black}{Therefore, these concerns and potential risks emphasise the need for further research to explore this implication more deeply.}

\subsection{\textcolor{black}{Location Privacy Attacks and Protection Mechanisms}}

\subsubsection{\textcolor{black}{Privacy Attack Methods on SNSs.}}

\textcolor{black}{Privacy attacks on SNSs involve unauthorised access, misuse, or exploitation of personal information shared by users. These attacks can lead to identity theft, unauthorised profiling, and other privacy breaches~\cite{zheleva2011privacy}, exploiting the extensive personal data on SNSs and posing significant risks to user privacy and security. Nowadays, various types of privacy attacks are prevalent on SNSs~\cite{beigi2020survey, cycberthreat2017, jain2021online, cerruto2022social}. }
\textcolor{black}{For example, \textit{Attribute Disclosure Attacks} can be used to infer sensitive personal attributes, such as political affiliations and age, without the user’s consent. For instance, ~\citet{zhong2015you} were able to deduce users' demographic information—such as age and gender—based on their location check-in data. More specific to location privacy, \textit{Location-Based Attacks} can be used to infer user movements and personal routines with users' exposed location profiles~\cite{li2016privacy}. Such information can be shared intentionally through tagging or unintentionally through metadata, which malicious users can then extract~\cite{cohen2022internet, GHAZINOUR2017267}. 
These attacks present serious threats to user privacy and safety. Therefore, educating users about potential risks, and promoting privacy-conscious behaviours are crucial steps in safeguarding personal information in the digital age.}

\begin{table*}[t]
\caption{Participant data, including demographics, used social media platforms, post frequency, and number of followers}
\centering
\scalebox{0.8}{\begin{tabular}{l l l l l l l}
\hline
\textbf{ID} & \textbf{Age} & \textbf{Gender} & \textbf{Education} & \textbf{Social Media Platforms} & \textbf{Post Frequency} & \textbf{Followers} \\
\hline
1 & 29 & F & Bachelor's & Instagram, LinkedIn & 3-7 times a week & 1001 - 5000 \\
2 & 27 & F & Bachelor's & Instagram, Twitter & 3-7 times a week & 501 - 1000  \\
3 & 31 & F & Bachelor's & Instagram, TikTok & More than 7 times a week & 1001 - 5000 \\
4 & 30 & F & Master's & Instagram, Facebook, TikTok, WeChat, Xiaohongshu & 1-2 times a week & 5000 - 10000 \\
5 & 27 & M & PhD & Instagram, WeChat & 1-2 times a week & 101 - 500 \\
6 & 27 & M & Master's & Instagram, Telegram & 1-2 times a week & 101 - 500 \\
7 & 22 & F & Bachelor's & Instagram, WeChat, Xiaohongshu & 3-7 times a week & 101 - 500 \\
8 & 22 & F & Bachelor's & Instagram, TikTok, Telegram & More than 7 times a week & 1001 - 5000 \\
9 & 25 & F & Master's & Instagram, TikTok, WeChat, Xiaohongshu & 3-7 times a week & 101 - 500 \\
10 & 27 & F & Bachelor's & Instagram, Facebook, Twitter, TikTok & 3-7 times a week & 1001 - 5000 \\
11 & 26 & M & Bachelor's & \makecell[{{l}}]{Instagram, Facebook, Twitter, Snapchat, \\ TikTok, Telegram, WeChat, Xiaohongshu} & 3-7 times a week & 101 - 500 \\

12 & 23 & F & Master's & Instagram, TikTok, WeChat, Xiaohongshu, Weibo & 3-7 times a week & 1001 - 5000 \\
13 & 20 & F & High school & Instagram, Facebook, Twitter, TikTok, Telegram, Xiaohongshu & 3-7 times a week & 501 - 1000 \\
14 & 37 & F & Master's & Instagram, Facebook & More than 7 times a week & 5000 - 10000 \\
15 & 19 & F & High school & Instagram, WeChat, Xiaohongshu & 1-2 times a week & 101 - 500 \\
16 & 22 & F & Bachelor's & Instagram, Facebook, WeChat, Xiaohongshu & 3-7 times a week & 1001 - 5000 \\
17 & 25 & F & Bachelor's & Instagram, Facebook, TikTok & 1-2 times a week & 1001 - 5000 \\
18 & 23 & F & Bachelor's & Instagram, Facebook, WeChat, Xiaohongshu &More than 7 times a week & 101 - 500 \\
19 & 28 & M & Master's & Instagram, Facebook, Twitter & 3-7 times a week & 501 - 1000 \\
\hline
\end{tabular}}
\label{tab:data}
\end{table*}

\subsubsection{\textcolor{black}{Emerging Risks of LLM Applications in Photo Location Privacy.}}

As online photo sharing becomes increasingly prevalent, users face significant privacy risks. \textcolor{black}{Recent AI techniques, for example, GeoSpy~\footnote{https://api.geospy.ai/}, excel in geolocation tasks by accurately determining locations from images and offering detailed reasoning for their analyses. However, these models often demand specialised technical expertise for setup, which limits their accessibility to users with the required knowledge.}
\textcolor{black}{Similarly, in another recent study conducted by~\citet{liu2024image}, research on
 large vision-language models revealed their alarming ability to extract geolocation data from images, even without explicit geographic training.}
\textcolor{black}{In a study conducted by ~\citet{wazzan2024comparing}, 60 participants were assigned to either traditional or LLM-based search engines as assistants for geolocation. Their study result showed that participants using traditional search more accurately predicted the location of the image compared to those using the LLM-based search. However, the researchers reported that over half of the participants using the LLM-based search reported challenges in formulating their queries.
Therefore, in our study, we integrate the query into the backend of the intervention app, which reduces technical barriers for participants and ensures a seamless experience.
Overall, the rise in popularity and accessibility of LLMs has made photo geolocation extraction more effortless than ever before, providing a new perspective for privacy attacks on SNSs beyond traditional photo tagging networks analysis~\cite{pesce2012privacy},} prompting the development of various strategies and technologies to mitigate these concerns.

\subsubsection{\textcolor{black}{Mechanisms for Protecting Privacy in Online Photo Sharing.}}
One of the primary mechanisms for protecting privacy in online photo sharing is the implementation of robust access control systems. These systems allow users to define who can view their photos, thereby enhancing privacy. For instance, platforms like Facebook enable users to select specific friends or groups for sharing content, which can significantly reduce the risk of unwanted exposure~\cite{malik2016impact}. \citet{toubiana2012photo} suggested that geolocation information should be used to automatically apply preset privacy preferences when a photo is taken, thereby reducing the likelihood of unintentional privacy breaches. Similarly, \citet{klemperer2012tag} argued that user-generated text tags can be employed to guide automated privacy and access controls for online images. 


In addition to access control, technological interventions such as image encryption and obfuscation techniques are critical in safeguarding user privacy. Techniques like Ciphertext-Policy Attribute-Based Encryption (CP-ABE) have been proposed to secure photo sharing by integrating access policies directly into the encryption process~\cite{yuan2017context}. Moreover, privacy-enhancing technologies such as face blurring and pixelation have been explored to obscure sensitive information within images before sharing~\cite{li2017effectiveness}. \textcolor{black}{Similarly, ~\citet{choi2017geo} suggest that including filters and cropping, already serve as natural geo-privacy protectors. In their experiments, up to 19\% of images whose location would otherwise be automatically predictable were no longer detectable after enhancement.}
However, one of the challenges is that this process can significantly diminish the visual appeal of the photos, which can deter people from adopting these privacy-enhancing techniques~\cite{hasan2020reducing}. \textcolor{black}{Building on this, \citet{hasan2018viewer} investigated 11 filters applied to obfuscate 20 different objects and attributes, exploring the trade-offs between image privacy and utility. They found that stronger filters generally offer greater privacy but often compromise user satisfaction and visual appeal. In contrast, weaker filters or context-specific obfuscation often achieve a better balance between privacy protection and visual quality.}


Lastly, the role of user awareness and education should not be underestimated. Research indicates that many users exhibit a lack of concern regarding privacy risks, often prioritising the social benefits of sharing over potential privacy infringements~\cite{wan2016advent}. This underscores the importance of developing educational interventions that inform users about the implications of their sharing behaviour and the available privacy protection mechanisms. Studies have shown that increasing user awareness can lead to more conscious sharing practices and a greater engagement with privacy settings~\cite{malik2016privacy}. 

Our study involves a longitudinal investigation using an app designed to \textcolor{black}{raise awareness of users of the risks that LLMs pose to their location privacy when sharing photos online. We provide valuable insights into the evolving landscape of location privacy challenges in the era of advanced AI technologies}. 

\section{METHOD}

\subsection{\textcolor{black}{Sampling and Recruitment}}
\textcolor{black}{
The recruitment process for this study began with a pre-screening survey promoted through our university's notice board. 
 Participants were selected based on the following criteria: (1) they were iOS users aged 18 years or older; (2) they posted photos on social media platforms at least once per week; and (3) they had a minimum of 100 followers on their photo-sharing social media accounts.  
By targeting active social media users with a minimum level of audience engagement, we ensure that participants are familiar with online photo-sharing practices and are more likely to have concerns about privacy breaches, but are not necessarily aware of the risks that LLMs pose on this topic.}

We recruited 19 participants (4 men, 15 women), \textcolor{black}{aged between 19 and 37 years, with a mean age of 25.8 years (SD = 4.3) for our user study.}
\textcolor{black}{Participants came from diverse educational and professional fields, spanning disciplines such as Computer Science and Information Technology, Nursing, Architecture, Business Administration, Marketing, Art and Design, Industrial Engineering, Economics, Medical Science, etc.}
The sample size is in line with privacy-related interview studies previously published at CHI~\cite{bourdoucen2024privacy,roemmich2023emotion} 
\textcolor{black}{and is consistent with the existing qualitative research literature on achieving thematic saturation~\cite{guest2006many}.}
\textcolor{black}{The gender distribution of our study is likely attributed to the fact that women generally express greater concern about sharing personal information on social networking sites compared to men~\cite{tifferet2019gender}, particularly regarding privacy risks related to location privacy~\cite{sun2015location}.}
Participant details are shown in Table~\ref{tab:data}. Participants were compensated with a \$65 gift voucher and our study was approved by our University’s Human Ethics Committee.


\subsection{\textcolor{black}{User Study} Procedure}

\begin{figure*}[htbp]
     \centering
     \includegraphics[width=1\linewidth]{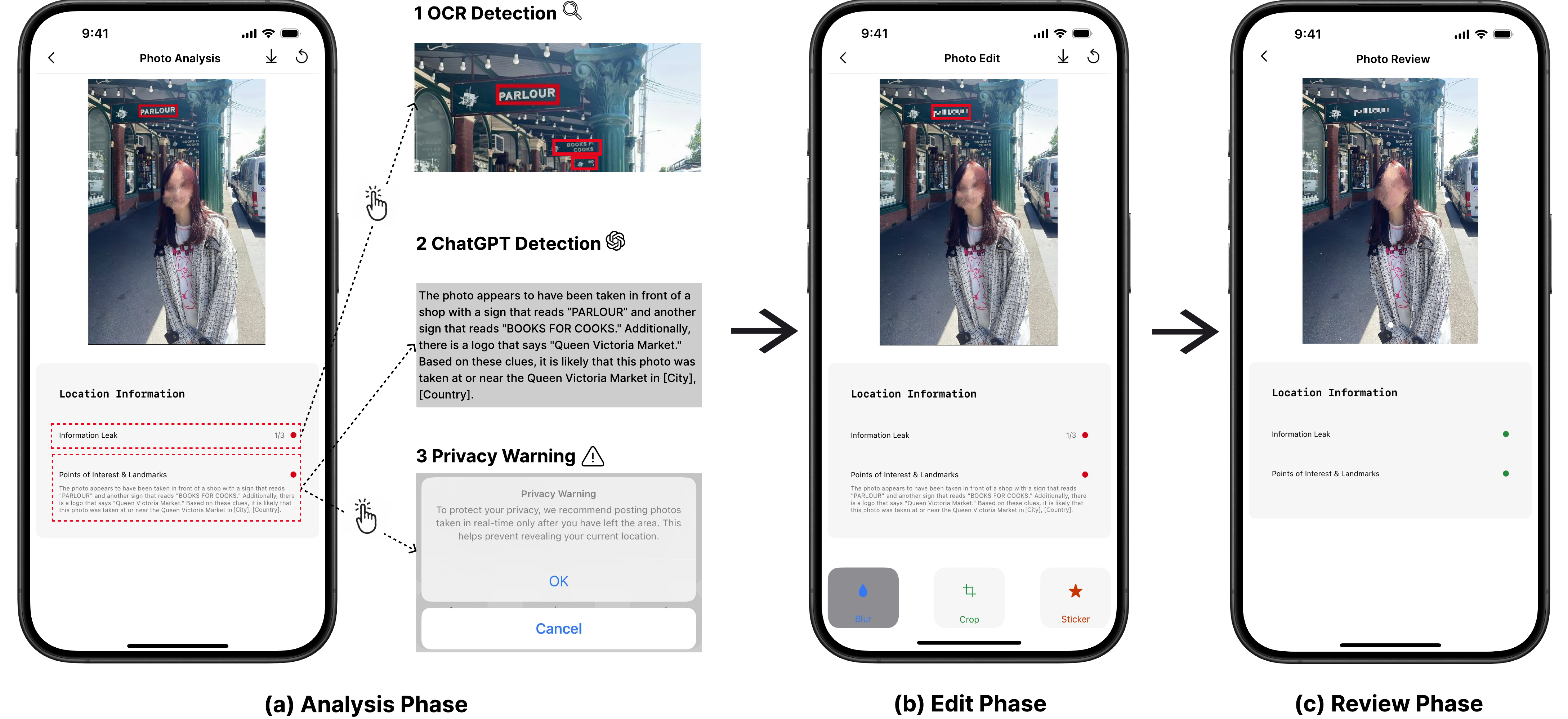}
     \caption{Different features of our application. (a) Analysis phase: where users get information of the location leaks in the photo they uploaded, (b) Edit phase: where users can mask location information, (c) Review phase: where users can verify the final photo and confirm that it no longer contains location leaks.}
     \label{Appflow}
    \Description{A horizontal workflow diagram showing three main phases of the app's interface, arranged from left to right. This includes an analysis phase where users get information about the location leaks in the photo they uploaded, an edit phase where users can mask location information, and a review phase where users can verify the final photo and confirm that it no longer contains location information leaks.}
\end{figure*}

The study was structured into three stages:

Stage 1: Initial In-Lab Session. Participants were invited to attend a brief, 15-minute session in the lab. During this session, the potential benefits and risks of the study were explained to them. \textcolor{black}{Participants were given the opportunity to ask questions and clarify any concerns before providing their informed consent by signing a consent form. After consenting,} participants installed our intervention app on their iPhones. The app's features were then demonstrated, and any questions were addressed. This session was designed to ensure that participants felt comfortable and clear about using the app.

Stage 2: Daily App Use for Two Weeks. Following the in-lab session, participants were asked to use the app daily for a period of 14 days. Each day, they assessed the privacy risks associated with the photos they took, and preferably, test the photos they intended to or had already shared on social media. 
\textcolor{black}{This phase allowed participants to use the app in real-world settings, enabling deeper reflection to inform the subsequent interviews effectively.}

Stage 3: In-Depth Interview. At the conclusion of the two-week period, participants took part in an \textcolor{black}{in-person} semi-structured interview lasting around one hour. This interview provided an opportunity to discuss their experiences with the app in detail. The questions focused on how the app influenced their privacy awareness  and intentions to change their location privacy practices, any challenges they encountered, and suggestions they had for future location privacy-preserving tools. The list of questions is shown in Appendix ~\ref{appendix_que}.

\subsection{App framework} 

We developed an iOS app aimed at raising users' awareness of location privacy leaks in photos taken by their smartphones. \textcolor{black}{Developing the app for iOS allowed us to streamline development due to greater platform uniformity when compared to Android devices, ensuring a more consistent research environment.}
\textcolor{black}{The app architecture follows a client-server model. On the client side, the app employs an MVVM (Model-View-ViewModel) pattern to manage the interface and data flow efficiently. The ViewModel facilitates data binding between the ViewController and the Model, ensuring real-time updates when users interact with the app. The server side consists of a RESTful API that handles requests from the client, a database client that processes queries, and a MySQL relational database, chosen for its efficiency in handling frequent metadata modifications and queries.}
The app operates as outlined in Figure~\ref{Appflow}. When the user opens the app, they are greeted by a home screen, where they are introduced to the app's functionalities.


Next, the \textbf{Analysis} phase, depicted in Figure~\ref{Appflow}(a), involves two key photo analysis features. First, the ``Information Leak'' step entails checking for textual information within the photo that can be used to infer the location where that photo was taken (e.g., street name, sign posts). We employed Optical Character Recognition (OCR) technology to detect texts in participant-uploaded images. Specifically, we used the pytesseract library~\footnote{https://pypi.org/project/pytesseract/}, which is an implementation of Google’s Tesseract-OCR Engine. Based on this analysis, the interface displays the specific number of text exposures\textcolor{black}{~(Figure~\ref{Appflow}(a)1)}. Following this, when the user presses the ``Information Leak'' feature, the OCR technology will progressively highlight the text by enclosing it in red rectangular boxes for mask processing in the following Edit phase. Second, in the ``Points of Interest and Landmarks'' feature, we used OpenAI's GPT-4 API to analyse the photo and identify the location using the prompt, \textit{``Can you identify the location where this photo was taken?''}\textcolor{black}{~(Figure~\ref{Appflow}(a)2)}. Due to the difficulty of masking points of interest and landmarks, such as famous bridges and known buildings, here we also provide a privacy warning message \textcolor{black}{\textit{-``\textcolor{black}{To protect your privacy, we recommend posting photos taken in real-time only after you have left the area. This helps prevent revealing your current location}''}}~(Figure~\ref{Appflow}(a)3). 
This warning pops up at the bottom part of the screen, suggesting users to consider sharing that photo only after leaving the area.

Following this, the \textbf{Edit} phase (Figure \ref{Appflow}(b)) provides three primary image manipulation functions: blurring, cropping, and sticker overlay. Upon selecting the blur function, the system automatically applies a mosaic effect to the area delineated by a red bounding box. As for the cropping function, users can freely drag and resize the box to crop the desired section. When users press the sticker button, the area within the red box is covered with an emoji selected by the user. In addition to the automatic masking operation, users can also manually perform these three editing actions on any area of the photo.

Finally, the \textbf{Review} phase~(Figure \ref{Appflow}(c)) involves a final check to ensure the accuracy of the entire process, allowing for any necessary corrections. The app provides an updated analysis of the photo to let the user know if their edits have successfully removed any location information leaks. \textcolor{black}{This feedback aims to empower users to make further adjustments if needed.} Once the user is satisfied with the photo, they can save it to their photo gallery. \textcolor{black}{Users can decide whether to share photos on SNSs, but we do not track subsequent sharing behaviour.}

\subsection{Data Collection \& Analysis}

We collected participants' usage data throughout the study, specifically recording each instance of a photo being analysed by the app, without storing the photo itself, and capturing data on whether it contained an information leak and whether points of interest \& landmarks were detected.

For the qualitative data, each interview lasted for around 45 minutes. 
\textcolor{black}{The interviews were audio-recorded and transcribed using iFLYTEK~\footnote{\url{https://www.iflytek.com/}}. To mitigate potential privacy risks, all participant data was anonymised before transcription, ensuring that no personally identifiable information was linked to the transcribed text.}
We employed reflexive thematic analysis, a six-phase process introduced by ~\citet{braun2021one}. 
In the first phase, we began by familiarising ourselves with the data. Interview transcripts were read to gain a thorough understanding of participants' responses, the context of \textcolor{black}{their reflections on photo-related location privacy, photo editing and sharing considerations,} and their experiences with the intervention app. 
In the second phase, initial codes were generated by identifying and labelling significant features of the data.
\textcolor{black}{Examples of codes developed in this phase include `awareness', `trust', `privacy education', etc.} \textcolor{black}{Then the first author gathered supporting sentences or paragraphs that contributed supporting codes.}
In the third phase, these codes were then organised into potential themes by grouping similar codes together. \textcolor{black}{For example, codes like `aesthetics concerns' and `photo editing preferences' were combined into a broader theme that captured participants' considerations of visual appeal and privacy; similarly, codes like `location-sharing decisions based on different contexts' and `comfort with location-sharing at a certain level of granularity' were used to develop the theme `importance of context \& granularity in online location privacy'.}
In the fourth phase, preliminary themes were reviewed and refined by the whole research team to ensure they accurately represented the data, with adjustments made to improve clarity and coherence. Each theme was clearly defined and named to reflect its content and significance, with definitions developed to articulate the essence of each theme and its relevance to the research questions. \textcolor{black}{At this point, three major themes emerged in this phase, which include awareness of the risks of LLM-powered location information leaks, the intention of behaviour change to safeguard online location privacy, and concerns and expectations for future location privacy-preserving technologies.}
In the fifth phase, the themes were further defined and categorised, consolidating into sub-themes. 

Finally, we wrote a detailed account of the themes, showcasing how they related to the overall research objectives. This analysis included detailed examples and quotes from the interviews.
\textcolor{black}{The first author led the data analysis while regularly reporting and discussing the process with the rest of the research team.}
We revisited phases 3 to 6 in an iterative cycle, refining the themes and ensuring a progressively recursive analysis and interpretation~\cite{trainor2021developing}.

\subsection{\textcolor{black}{Ethical Considerations}}
\textcolor{black}{
Given the privacy-oriented nature of this research, which aims to help users safeguard their location information leaks from photos, we took careful steps to ensure participants were fully informed about the study's procedures.
Before participating, all participants received an email containing the Consent Form and Plain Language Statement, which outlined potential risks and benefits, and the tasks participants would be asked to complete.
To emphasise transparency, during Stage 1's initial in-lab session, participants were given time to review the consent form again, ask any questions they had regarding the terms and concerns related to data usage practices, and sign the form physically. They were also verbally briefed by researchers and informed that the database would not collect any photos uploaded through the app or their GPS location data. Participants were further informed that LLM outputs and app usage data would be recorded.
}

\textcolor{black}{
An important point to raise is the decision to use OpenAI's API for image analysis, as it may appear to contradict the study's focus on privacy protection. However, OpenAI explicitly states that ``We do not train our models on inputs and outputs through our API''~\footnote{https://platform.openai.com/docs/concepts}. This policy mitigates the privacy concerns associated with sharing data with third parties. We acknowledge that leveraging a locally hosted, open-source model~\cite{xu2024device} for image analysis would offer an alternative that better aligns with privacy principles by eliminating reliance on external services. Future efforts could focus on integrating such models~\cite{chu2024VLM,zhang2024enabling} into an upcoming app to further increase user privacy. However, due to concerns about model accuracy, implementation complexity, and to reduce the burden on participants~\footnote{https://www.datacamp.com/blog/the-pros-and-cons-of-using-llm-in-the-cloud-versus-running-llm-locally\#rdl}, deploying and maintaining a locally hosted model on participants' devices was deemed impractical. As a result, we opted to use OpenAI's large language model, thoroughly informing the participants of our approach, recognising the trade-offs but prioritising practical feasibility.}

\section{Findings}

\textcolor{black}{During the two-week study period, a total of 682 photos were analysed, with an average of 35.9 photos per participant. Of these, 54.7\% of photos were identified as having potential location leaks, while 45.3\% were detected as having no location leaks. Within the subset of photos flagged for potential location leaks, 51.5\% were identified as only containing information leaks (textual information), 21.4\% were identified as only containing identifiable points of interests and landmarks, and 27.1\% contained both types of location information.}

In this section, we present the qualitative findings from our study, offering insights into \textcolor{black}{awareness of the risks of LLM-powered location information leaks and intention of behaviour change regarding online location privacy.}
In addition, we analyse their perspectives on future location privacy technologies, including potential measures to improve trust in privacy protection tools.

\subsection{\textcolor{black}{Awareness of the Risks of LLM-powered Location Information Leaks (RQ1)}}

\subsubsection{\textbf{Surprise over the App's Capabilities}}

Participants \textcolor{black}{(N=14) }
universally expressed surprise at how advanced technology could accurately detect their location, even when the location cues in the photos were minimal or subtle. This level of sophistication far exceeded their expectations, highlighting a gap in user awareness about the capabilities of modern technology in location detection.

\begin{figure*}[htbp]
     \centering
     \includegraphics[width=0.85\linewidth]{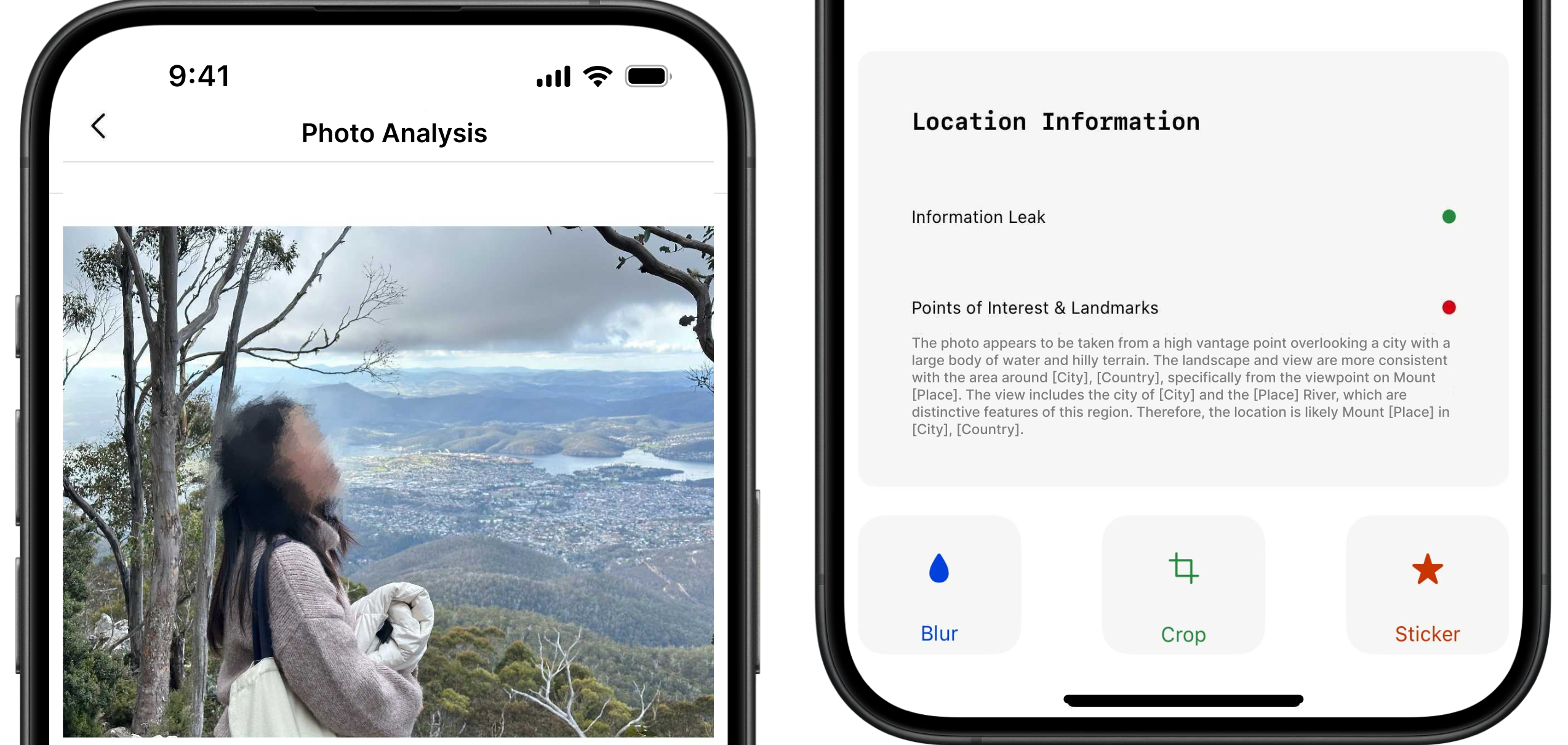}
     \caption{\textcolor{black}{(Left) A photo shared by a participant indicating she is on a mountain landscape; (Right) Detected points of interest and landmarks from this photo. (The participant consented to the use of this photo in the paper.)}}
     \label{paricipant_case}
     \Description{The two figures illustrate an example of the app’s photo analysis interface. The left figure shows a participant standing on a mountain landscape, facing away from the camera, with a city, water body, and hilly terrain in the distant background. The right figure presents the location information extracted from the photo. The interface includes two sections: 'Information Leak,' which was not detected in this example, and 'Points of Interest \& Landmarks,' which was detected and provides a textual description of identifiable location details from the image. Below this description, three privacy-preserving actions—'Blur,' 'Crop,' and 'Sticker'—are available, each represented by an icon.}
  \end{figure*}

For example, one participant described their amazement at the app's ability to identify their precise location, despite the similarity of the surroundings with other places~\textcolor{black}{shown in Figure~\ref{paricipant_case}}, stating,\textit{``I thought that all mountains should look the same. But then the app recognised I was in [City]. It was unbelievable! So, if an app can analyse and provide such specific location information, I think it would be very useful''}~(P4). Similarly, another participant recounted another experience where they uploaded a photo taken at a lecture hall in her university. The participant shared, \textit{``There was just a tiny logo in the corner of the photo---just our school's emblem without any text saying [redacted] University. Yet the app still identified it, which left me shocked''}~(P7).
\textcolor{black}{Another participant added, \textit{``Over time, I believe the AI will definitely improve further, becoming more accurate in location detection and capable of detecting even finer details''}~(P19).}

These reflections highlight the growing sophistication of AI and other technologies in detecting precise location information from subtle cues in photos, which has significant implications for user privacy. Participants were previously unaware of the risks, particularly the extent to which their location could be inferred.





\subsubsection{\textcolor{black}{\textbf{Indirect Location Information Leakage from Photos}}}

\textcolor{black}{After a two-week period of using our app, participants (N=4) also reflected on additional instances of potential indirect location information leakage from photos. For example, one participant shared concerns about screenshots and how hidden timestamps could inadvertently reveal their location, particularly if malicious actors leverage advanced AI technologies (P15).} 

\textcolor{black}{Similarly, another participant also pointed out that the carrier name displayed in the status bar of a screenshot might inadvertently disclose their current country, particularly while travelling internationally. \textit{``When I try to upload a screenshot for analysis, the app's information leak part detects the carrier name and highlights it, reminding me that the carrier's name can reveal my current country. Because I use a different carrier when I'm in my home country''}~(P6).}


\textcolor{black}{These examples highlight a heightened awareness of potential location privacy leaks in photos that could be easily detected by an LLM, especially when their location or travel status is sensitive.}

\subsection{ \textcolor{black}{Intention of Behaviour Change of Online Location Privacy (RQ2)} }

\subsubsection{\textbf{\textcolor{black}{Shifts in Photo-Taking and Sharing Practices}}}
\label{trust}
Participants 
\textcolor{black}{(N=10) expressed an intention to adapt} their photo-taking habits and social media sharing \textcolor{black}{after using this intervention}.
\textcolor{black}{This intention is reflected in their increased mindfulness about capturing identifiable information in photos.}
For example, one participant shared, \textit{``Sometimes I am more mindful of text. If I see a street name or a bus stop, I move my camera away from it so that I wouldn’t capture that''}~(P17). Similarly, another participant emphasised the importance of being cautious with potentially sensitive details \textcolor{black}{they may not have been aware of before} in the photo: \textit{``When I take photos, I will evaluate which parts of the image could expose sensitive information. \textcolor{black}{Then, I upload the image to the app to compare if I was correct or if there is something I missed that the app detected}''}~(P18), which indicates a proactive approach to protecting privacy.

Moreover, the hesitation to share photos is reflected in the experiences of several participants \textcolor{black}{(N=7)}, 
who \textcolor{black}{will reconsider posting photos} online, \textcolor{black}{signaling an increased awareness of potential risks. } As one noted, \textit{``Definitely, you would think twice about posting''}~(P17). This awareness and adaptation illustrate the growing influence of privacy concerns on everyday behaviours, particularly in the context of photo-sharing on social media platforms. \textcolor{black}{Interestingly, some participants (N=3) felt they would be more comfortable sharing more photos with the assistance of our app. They thought the app worked as a `moderator' (P19), reducing their concerns when sharing photos online. As one participant expressed: \textit{``I think this app can help protect my privacy when I share things, so I feel my likelihood to share has increased''}~(P4).}

The \textcolor{black}{intended} shift in behaviour mirrors the broader trust issues participants have with institutions \textcolor{black}{(N=14)} , 
especially large tech companies like Meta, \textcolor{black}{as one participant stated \textit{``Due to their commercial nature and past scandals involving the leakage of user privacy or the use of users' location data for training models, targeting ads, etc''}~(P11).} Such companies were seen as more likely to misuse data or be vulnerable to breaches. Trust in privacy protection is crucial, and participants showed a tendency to be more cautious when they believed their location data might be misused or mishandled, reinforcing the need for better privacy management in digital spaces.

\subsubsection{\textbf{Importance of Context \textcolor{black}{\& Granularity }in Online Location Privacy}}
\label{context}
\textcolor{black}{While our app surprised participants by showcasing AI's ability to quickly and accurately detect their location from photos using subtle cues, } the context still played a key role in users' decisions, with some locations seen as less sensitive despite heightened awareness.
When discussing the intention to tag locations on social media, some participants \textcolor{black}{(N=3)} 
expressed that they are fully aware of the implications of uploading photos and are often less concerned about their online location privacy. One participant mentioned, \textit{``I’m totally aware of what I post on social media, whether it’s personal information or just a casual post. Sometimes on Instagram, I’ll upload a picture of something I ate at a restaurant. I know we have to be cautious, but for me, it’s not that serious since it’s not personal or work-related information''}~(P1). This reflects a mindset where some users feel comfortable sharing their location details in specific contexts, especially when the content is seen as trivial.

A few participants \textcolor{black}{(N=5)}, particularly when travelling or dining at restaurants, expressed a preference for tagging their location. This behaviour often stemmed from the desire to document their experiences for personal memories or to share with friends. As one participant explained, \textit{``Posting photos and tagging the location helps me keep track of my life’s events, making it more organised and easier to remember''}~(P7).

However, while some participants \textcolor{black}{(N=3)} felt comfortable tagging broader areas such as countries or cities, they were less willing to share exact locations online, preferring to maintain a degree of privacy. As one participant stated: \textit{``I think it would be useful if you could anonymise the exact location, but in terms of the broad location, like which city or country I'm in, I think that's fine for me''}~(P19).\textcolor{black}{This participant further elaborated on the importance of \textit{proximity} and \textit{timing} on location sharing, noting that if the distance between the current location and the location in the photo is short, it could pose a potential threat.}
This balance between sharing and protecting location privacy underscores the nuanced way in which users manage their online presence and location data.






\subsubsection{\textbf{Attitude Towards Privacy Warnings}}
\label{privaywarning}
We provide a location privacy warning in the app intervention when points of interest and landmarks are detected, as shown in Figure~\ref{Appflow}a(3).
However, participants \textcolor{black}{(N=9)}
expressed mixed attitudes towards privacy warnings. Some participants emphasised the need for privacy warnings to appear repeatedly to remind them. As one participant remarked, \textit{``There are many privacy-related one-liners, and even if you bombard people with them, you have to repeat them again and again''}~(P14). This highlights the challenge of maintaining user attention when warnings are frequent and repetitive.

Another participants shared a different sentiment, recognising the importance of warnings but also noting the potential for desensitisation over time. They stated, \textit{``Warnings are useful, but after seeing them too many times, they can become numbing, and I may not feel as alert''}~(P18). This participant further explained that while they might ignore low-risk warnings, such as those related to non-sensitive information like the layout of a home’s interior, they would pay closer attention to higher-risk warnings, especially those involving identifiable information like a home address. \textcolor{black}{P19 suggested that showing the potential consequences of location leakage in privacy warnings would increase user awareness and attention.}
\textcolor{black}{In addition, P14 further suggested that they would prefer having different AI-based warnings and tips tailored to various scenarios. This approach could also be an effective way to educate users.}
These insights suggest that while users appreciate privacy alerts, the frequency and context of the warnings influence their overall effectiveness in prompting protective actions.

\subsection{\textcolor{black}{Concerns and Expectations for} Future Location Privacy Technologies \textcolor{black}{(RQ3)}}

\subsubsection{\textbf{Balancing Visual Appeal \& Privacy Considerations}}

Participants \textcolor{black}{(N=10)}
expressed varying degrees of concern over the balance between maintaining the visual appeal of the photo and ensuring privacy. For some, the idea of editing photos shifted from focusing on how visually appealing the content was to considering more potential privacy risks. One participant remarked, \textit{``I would just focus on what I'm sharing in terms of the looks, like it looks good, looks interesting. But now I would rather share less potentially or share photos that are less pleasing, less aesthetic, but more secure''}~(P6). This illustrates a growing awareness of privacy issues, leading to decisions that prioritise security over aesthetics.

\textcolor{black}{Other participants (N=7) noted how they became more vigilant about subtle details that could inadvertently reveal sensitive information while editing photos. However, one participant shared the difficulty they faced:\textit{``Sometimes, it's challenging to edit photos to look good while also ensuring that my location information remains hidden, especially when the majority of the photo is identifiable place''}~(P15). This highlights the need for photo editing tools that incorporate privacy-preserving features, allowing users to easily conceal sensitive details like location without sacrificing much visual quality.}

Another participant described how she used editing tools to both enhance aesthetics and protect privacy by covering sensitive information. \textit{``I don’t usually blur or cover things up, but sometimes on Instagram, I write text over the photo. The app lets you add a background to the text, so I’ll use that to block out anything I don’t want to show''}~(P16). This approach demonstrates expectations and effort to balance visual appeal with privacy concerns by creatively using aesthetic tools to safeguard personal information.

These reflections highlight a trade-off many users are now making—sacrificing some aesthetic quality in exchange for enhanced privacy, as they become more conscious of the risks associated with sharing personal information through photos.

\subsubsection{\textbf{Real-Time Detection over Post-Hoc Solutions}}

A few participants \textcolor{black}{(N=3)} expressed a strong preference for real-time \textcolor{black}{photo} privacy detection over post-hoc solutions, viewing it as more effective in preventing potential privacy risks before they occur. One participant noted, \textit{``Real-time detection is likely to be more helpful. \textcolor{black}{Since it is difficult to remove location information from certain images through editing}''}~(P16). They emphasised that if an app could analyse a photo as it is being taken and provide immediate feedback, such as highlighting identifiable landmarks or sensitive information, it would allow users to adjust the photo before sharing. This proactive approach would reduce the need for post-editing and offer users greater control over their privacy in the moment, enhancing their trust in the tool’s ability to safeguard their personal information.



\subsubsection{\textbf{Privacy Education}}

Participants \textcolor{black}{(N=6)} highlighted the critical role of privacy education in enabling users to make informed choices regarding their online activities. One participant, who has lived in five different countries, observed the influence of education disparities on privacy awareness. Urban dwellers, particularly those with higher levels of education, tend to exercise greater caution in sharing personal information online. Conversely, individuals from rural areas may lack awareness of potential location privacy risks, leading to more permissive sharing behaviours. As they expressed:
\textit{``I think a lot of people who grew up in cities which have more education [...], or [are more] literate, are more aware of their privacy concerns''}~(P10).

Furthermore, a couple of  participants suggested that social media platforms should adopt a more proactive approach to educating users about privacy. They recommended integrating educational content within the platforms, such as tutorials and community forums, to raise awareness about the importance of safeguarding personal \textcolor{black}{location} information. By offering practical guidance and real-life examples, 
\textcolor{black}{it can demonstrate how photos can unintentionally reveal location information, leading to significant privacy risks and potential consequences such as personal security threats, unwanted tracking, or targeted marketing efforts. These real-life scenarios can help users better grasp the potential dangers, empowering them to more effectively manage their location privacy online.}
As another participant noted: \textit{``I think there should be tutorials or maybe a community where people can share tips on how to protect their privacy. It would really help users become more aware''}~(P11).
These perspectives underscore the need for social media platforms to incorporate educational resources that empower users to navigate privacy challenges effectively. Providing such tools could lead to more informed and cautious online behaviour, ultimately enhancing privacy protection.

Another participant further highlighted the importance of self-education, noting that many users skip over consent forms and terms of service agreements when using apps. They remarked:
\textit{``Because like I feel sometimes when we use apps, they already have like a consent letter and it is just like the people who don't read the terms and conditions. So I feel it's like to the user itself on how they can educate themselves about social media and posting''}~(P1).

These perspectives underscore the need for social media platforms to incorporate educational resources that empower users to navigate \textcolor{black}{location} privacy challenges effectively. Providing such tools could lead to more informed and cautious online behaviour, ultimately enhancing \textcolor{black}{location} privacy protection.


\subsubsection{\textbf{AI-Assisted Decision-Making for Location Privacy}}

Several participants \textcolor{black}{(N=5)} expressed varying levels of concern regarding location privacy when using AI to assist with photo-sharing decisions or photo-editing suggestions. 
One participant highlighted the potential usefulness of AI in curating a refined set of photos for review, particularly with respect to balancing aesthetics and privacy. \textcolor{black}{ 
As one participant noted: \textit{``AI can help remove some location privacy-related objects more naturally''}~(P8).
Similarly, another participant remarked: \textit{``But if it helps to select out of this bunch, which option is the most location information-safe, and if you can add to the rubric which is the most aesthetic and looks nice. I think I'd definitely enjoy that''}~(P17). They further elaborated on their perspective, emphasising the complementary role of AI in supporting, rather than replacing human decision-making: } \textit{``I won't say it's better than human, but I think AI can do a good job at least. For example, if AI can refine a bunch of photos, and then you pick from there, that would be very useful. But the final decision of what I post will still be up to myself''}~(P17). This reflects that participants saw AI as a helpful tool in filtering and editing options but retained the final decision-making authority due to personal preferences and the nuances that AI might miss, such as sentimental value or specific contextual details in the image.

Another participant highlighted the challenges AI faces in identifying privacy-sensitive elements in photos, emphasising that individuals have different degrees of concern about privacy. For example, \textit{``AI can’t know some things that I don’t want others to see... like if my friend's birthday gift is on the table. When I see it myself, I know to put a sticker over it, but AI wouldn't know that''}~(P8). 
\textcolor{black}{This highlights that users are optimistic about AI's potential to assist with privacy-related decisions but remain cautious, recognising that }it may not fully replace human judgment, especially in complex, context-dependent scenarios.

Therefore, AI-generated solutions for location privacy in photos are seen as helpful but not infallible. While participants appreciated the efficiency of AI in sorting and identifying potentially sensitive images, they consistently emphasised the importance of human oversight in the final decision-making process.

\subsubsection{\textbf{Transparency in Location Privacy-Preserving Technologies}}

\textcolor{black}{Several participants (N=6) highlighted the critical need for transparency in location data practices to mitigate concerns about location privacy breaches. Several participants expressed unease about the invisibility of data handling processes specific to location information, such as the mechanisms behind location tracking, the criteria for sharing location data with third parties, and the duration for which such data is retained. This lack of clarity created a pervasive sense of mistrust.}
\textcolor{black}{For example, one participant noted the difficulty in understanding how their location data might be extracted without their explicit awareness. They expressed concerns about the multiple ways location information could be obtained, \textit{``If location permissions are not enabled, I am wondering if these SNSs can still obtain my geographic location through other means, such as my carrier or inferring it from the locations of my followers''}~(P6). This underscores the general skepticism users have regarding online location privacy and emphasises the need for greater clarity between users and platforms concerning location data extraction processes.} 

\textcolor{black}{Another participant raised the issue of ambiguity in consent mechanisms for location data sharing, stating that apps often bundle location permissions with general data terms, making it challenging for users to discern what they are agreeing to. As he pointed out: \textit{``Some SNSs display a pop-up explaining the specific purpose when they require me to enable location permissions, but most SNSs, as far as I can remember, describe it in very general terms''}~(P19). This led to calls for interfaces that clearly delineate the purposes of location data collection and provide real-time feedback on how the data is being used.}

\textcolor{black}{Even with the introduction of interventions to assist users in identifying potential location privacy leaks, participants emphasised that, if such features were embedded directly within SNSs, it would still be crucial to provide clear explanations of how the analysed data would be utilised by the platforms. One participant stated, \textit{``I want to know more information, like what the model knows about my photo and what it doesn’t know—something like that. It’s good to be transparent. Also, how will the analysed location data be used''}~(P3).}



Therefore, increased transparency around data handling and privacy measures could alleviate user concerns, creating a more trustworthy user experience.

\subsubsection{\textcolor{black}{\textbf{System or SNS Integration}}}
\label{On-Device}
\textcolor{black}{Participants (N=4) expressed that the app's perceived trustworthiness significantly increases if it can be integrated within trusted platforms like iOS. 
For example, one participant emphasised the benefit of on-device processing, noting, \textit{``Cloud servers may offer better performance, but the advantage of local processing is that users can ensure their privacy. If the app doesn’t connect to the internet, it's definitely more trustworthy, especially if it's deeply integrated into the system, like being part of Apple's Photos''}~(P11). At the same time, this participant highlighted that when a photo is shared via AirDrop, iOS's sensitive content warning feature detects and flags images or videos containing nudity. They suggested that a similar approach could be implemented for location privacy, where Apple could apply the same detection capabilities to safeguard users' geographic information.}

\textcolor{black}{Social media integration was also mentioned as a potential feature, allowing for real-time privacy protection directly within platforms, reducing the need for users to switch between apps.
As one participant said,  \textit{``If it's directly embedded in social media, it would be very convenient. It could help me check potential location exposure before I post, acting like a `filter'''}~(P12).
} 

\section{Discussion}


\subsection{Location Privacy: Awareness, Autonomy, and Education}
Our intervention app prototype allowed participants to test the potential location privacy risks associated with the photos they planned to share on social media or have already shared on social media before, letting us understand their attitudes and opinions on this privacy-preserving technology.
Previous work suggests that it is essential to highlight that when users upload photos to social media, third parties can extract location data embedded within these images, leading to unintended privacy leaks~\cite{li2015deciphering}. With the increasing sophistication of AI technologies, particularly \textcolor{black}{the widespread adoption of LLMs}, our study reveals that users are surprised by \textcolor{black}{these AI tools' ability to extract location information from visual elements in photos and} how accurately it can identify and infer exact location addresses from seemingly innocuous photos. This unexpected capability heightened \textcolor{black}{the need to raise} awareness of potential location privacy risks on SNSs.

\textcolor{black}{Previous studies highlighted that users regard location metadata as the most sensitive information embedded in photos~\cite{henne2013awareness}.  Growing awareness among platforms and individuals has led to an increased focus on encrypting photo metadata before sharing to protect privacy. A decade later, the rise of LLMs introduces new challenges for location-related privacy of photos. 
In response to these evolving challenges, we propose the development of \textbf{tools and mechanisms that grant users full autonomy to control their location data}.
For instance, users could benefit from an AI-powered ``privacy scan'' feature that analyse a photo before sharing, highlighting potential location indicators. 
Additionally, tools could allow users to customise the level of privacy protection based on the context or intended audience. A dynamic preview feature could demonstrate how these adjustments alter the likelihood of someone deducing the location from the photo, enhancing users' awareness of potential risks.}
By providing users with better insights into these risks, they can make more informed decisions about their online location privacy, rather than relying on the limited options currently available on mainstream platforms.

In addition, there is a pressing need to enhance users' privacy literacy on social media platforms, as highlighted by recent research from~\citet{choi2023privacy}.  \textcolor{black}{However, several challenges remain, possibly due to users' behavioural inertia, conflicts of commercial interests, and insufficient educational formats and media~\cite{finkelhor2021teaching}.}
Furthermore, current consent mechanisms (e.g., terms and conditions agreement) are insufficient in this regard~\cite{asthana2024know}. These mechanisms often fail to adequately inform users about the complexities of privacy risks, leaving them unaware of the potential consequences of their data-sharing decisions. 
Therefore, we need to rethink current consent frameworks to incorporate more effective educational elements that actively engage users in better understanding and managing their online location privacy~\cite{feng2021design}. 
\textcolor{black}{For example, platforms could integrate ``privacy tips'' by showcasing real-life examples of unintended location leaks from photos. By presenting relatable scenarios and highlighting the consequences of these privacy breaches, these tips can empower users to recognise potential risks in their own photo-sharing practices. Additionally, community-driven learning initiatives could engage users in identifying location risks within sample photos, fostering collaborative awareness in an interactive and relatable way. For instance, online communities or forums hosted by platforms could encourage users to share anonymised examples of photos with potential location risks and discuss strategies to mitigate them.}


\subsection{Enhancing Trust and Transparency in Privacy Protection Interventions}
Building trust in privacy protection technologies is essential to encourage their adoption. Our findings \textcolor{black}{(Section~\ref{trust})} highlight a significant distrust among participants regarding the use of location data by large corporations, particularly when there is a lack of transparency in how the data is collected and used. Users are more likely to trust privacy tools that operate locally on their devices, without the need for cloud-based processing, as this reduces concerns about data breaches and misuse. To foster trust, platforms should prioritise transparency. For example, providing clear, concise explanations of how location data is used, stored, and shared, along with giving users the ability to review and delete their \textcolor{black}{historical location} data at any time, will help build confidence in the technology. Moreover, third-party  \textcolor{black}{and government} audits or certifications could further reassure users that privacy practices adhere to strict standards \textcolor{black}{and corresponding laws}.

In addition, our study reveals an important ethical challenge associated with privacy-preserving technologies: while designed to safeguard users' privacy, these tools also present the potential for misuse. Participants raised concerns about their data being used to train models, which aligns with growing apprehension in the research community about the ethical implications of data being repurposed for AI development~\cite{kapania2024m}. To address these concerns, we recommend integrating local LLM models directly into the system, rather than relying on cloud-based or external services. Embedding LLMs locally on users' devices would minimise the risks associated with data sharing and external model training, offering users greater control over their data while ensuring privacy is maintained within the system architecture~\cite{zhang2024enabling}. Furthermore, incorporating the models directly into the system, rather than as a separate app, would enhance both usability and security by streamlining the privacy protection process. This approach would also ensure that users are well-informed about potential privacy risks before any photos are published online, allowing them to make more informed sharing decisions.

\subsection{Towards Effective Location Privacy-Preserving Mechanisms}

Our study's findings also highlight several important considerations for enhancing location privacy for photo-sharing on social media platforms. First, participants expressed a desire for real-time feedback that not only identifies potential location privacy risks but also educates them on the consequences of sharing specific photos. \textbf{This suggests the need for privacy mechanisms that are more proactive, such as live detection of sensitive location cues as photos are being taken}, rather than relying on edits. By doing this users can be made aware of potential risks in real time, enabling more informed decision-making, \textcolor{black}{which was seen as crucial given the capabilities of widely available LLMs}.

Furthermore, \textbf{designing privacy technologies that balance visual or aesthetic appeal and security is essential}. Several participants emphasised the tension between maintaining photo quality and concealing location-revealing elements. 
To achieve this, future systems should offer flexible editing options that preserve the visual integrity of images while safeguarding privacy. For example, AI-driven solutions could suggest subtle modifications, such as intelligently blurring backgrounds or removing identifying landmarks \textcolor{black}{more naturally}, without significantly altering the visual appeal of the photo.

\textcolor{black}{In addition, from the examples presented, Figure~\ref{Appflow} shows a case where the location becomes undetectable when textual cues are removed, effectively safeguarding privacy without significant aesthetic compromise. In contrast, Figure~\ref{paricipant_case} showcases a scenario involving landmark-like visuals. Here, extensive background blurring or removal can significantly diminish the photo's visual appeal. This highlights a challenging trade-off between preserving aesthetics and ensuring privacy. In such cases, emphasising the role of privacy warnings becomes paramount. By alerting users to potential location privacy risks before sharing content online, these warnings serve as a vital intervention. From our findings in section~\ref{privaywarning}, these insights underline the importance of context-sensitive and well-calibrated privacy interventions. While users value privacy alerts, the frequency, specificity, and clarity of warnings greatly influence their effectiveness in motivating protective actions. By striking a balance between proactive warnings and avoiding user fatigue, privacy-preserving interventions can empower users to make more informed decisions when sharing content online.
}

Lastly, our study reveals that proximity to the identified location and post timing \textcolor{black}{(Section~\ref{context})} are both critical factors influencing users’ location privacy concerns when sharing photos on social media. Participants highlighted the risks associated with sharing locations that are close to their current position, as this could potentially enable others to track their movements. This observation is consistent with previous research in HCI, which also emphasises the significance of proximity in location privacy~\cite{song2020m}. Additionally, post timing can serve as an indirect cue to infer a user's location area. These findings highlight the need to integrate considerations of proximity and timing into location privacy-protecting tools. By implementing features that delay location-sharing posts or proactively alert users to potential privacy risks by considering both the proximity of posted locations and the timing of the posts, users can mitigate the risks associated with real-time location tracking.




\subsection{\textcolor{black}{Limitations and Future Work}}

\textcolor{black}{In this section, we discuss our study's limitations and suggest potential avenues for future work on this topic. First, we developed an iOS app aimed at raising users' awareness of location privacy leaks in photos taken by their smartphones. However, we acknowledge that iOS users may differ from Android users in terms of privacy attitudes or technology use~\cite{benenson2013android}. Future work could extend this research by exploring potential differences in user interactions and privacy perceptions across mobile ecosystems.}

\textcolor{black}{Second, our sample was predominantly composed of individuals in their twenties (mean age = 25.8 years), with only three participants in their thirties. As such, our findings may not reflect the privacy needs and usage patterns of older adults, who might have different privacy preferences or face distinct challenges when using privacy-protection tools. Future research should aim to recruit a more age-diverse sample, including middle-aged and older participants, to understand how privacy intervention effectiveness and app usage patterns vary across different age groups.}
\textcolor{black}{Our study sample also exhibited a notable gender imbalance, with a predominance of female participants (see Table~\ref{tab:data}). This skew was also present in the recruitment process, where more females expressed interest in participating when the study was advertised. The predominance of female participants in our study can be attributed to the fact that women generally express higher concerns regarding online personal information sharing compared to men~\cite{tifferet2019gender}, potentially increasing their motivation to participate in privacy-focused studies. Future research should consider expanding to a more gender-balanced sample to enable comparative analyses. This would help identify potential gender-specific differences in privacy tool adoption, usage patterns, and intervention effectiveness, ultimately leading to more inclusive and comprehensive privacy solutions.}


Third, while our study spanned two weeks, this duration \textcolor{black}{is unlikely to lead to actual behavioural changes. Further research should consider extending the study period and capture actual changes in long-term user behaviours regarding location privacy practices instead of only intended changes}. Finally, our app offered limited functionality with regards to editing photos (blur, crop, stickers). This was deemed sufficient for the purposes of our study, but future work could explore a wider array of editing features.

\section{Conclusion}
\textcolor{black}{Our study provides an in-depth exploration of LLM-powered location privacy} risks associated with photo sharing on social media, offering valuable insights into how a location privacy-preserving intervention impacts users' awareness and intended behavioural changes. We found that users are surprised about the current capabilities of technologies for location detection and become more cautious in their intended sharing behaviour when they realise the potential for location leaks.
Moreover, our app proved to be valuable for users, providing them with a greater sense of control over their personal information. 
We emphasise the significance of users' awareness, autonomy, and education, alongside the need for trust and transparency from the platform's perspective.
Finally, our study yields important design implications for future location privacy-preserving tools. These insights help to better understand location privacy on social media, paving the way for more user-centred and effective privacy-preserving human-computer interactions.

\bibliographystyle{ACM-Reference-Format}
\bibliography{bibfile}

\newpage

\appendix

\section{Interview Questions}
\label{appendix_que}

The list of questions was prepared for each interview participant.
\subsection{Usage Experience}
\begin{enumerate}[label=Q\arabic*., font=\itshape, itemsep=0.5em, leftmargin=*]
    \item How often did you use the app during the two-week period?
    \item Can you walk me through a typical scenario when you used the app? 
    \item In what situations do you find this app helpful, and in what situations do you find it unnecessary?
\end{enumerate}

\subsection{\textcolor{black}{Awareness and} Behaviour Change}
\begin{enumerate}[resume, label=Q\arabic*., wide=0pt, leftmargin=*, font=\itshape]
    \item Photo-taking Behaviour
    \begin{enumerate}[label=Q\arabic{enumi}.\arabic*., wide=0pt, leftmargin=*, font=\itshape]
        \item Since you started using the app, have you noticed any changes in how you take photos? If so, can you describe these changes?
        \item Have you become more cautious about the types of photos you take or share? Can you give an example?
    \end{enumerate}
    \item Share Intention
    \begin{enumerate}[label=Q\arabic{enumi}.\arabic*., wide=0pt, leftmargin=*, font=\itshape]
        \item Did the app influence your intention/willingness to share photos on social media?
        \item Can you give an example of a situation where you decided not to share a photo after using the app?
    \end{enumerate}
    \item Edit Behaviour
    \begin{enumerate}[label=Q\arabic{enumi}.\arabic*., wide=0pt, leftmargin=*, font=\itshape]
        \item Did the app make you more likely to review or edit the your photos before sharing them?
        \item How would you prefer to edit your photos before uploading them on social media?
    \end{enumerate}
    \item Social Media Setting
    \begin{enumerate}[label=Q\arabic{enumi}.\arabic*., wide=0pt, leftmargin=*, font=\itshape]
    \item Have you made any changes to your social media habits or settings due to concerns raised by the app?
    \item Have you taken any additional steps to protect your privacy online since you started using the app?
    \end{enumerate}
\end{enumerate}

\subsection{App Design}
\begin{enumerate}[resume, label=Q\arabic*., wide=0pt, leftmargin=*, font=\itshape]
     \item What major challenges or difficulties have you encountered while using this app?
     \item What additional features would you like to see included in this app to make the app more useful?
     \item Do you have any suggestions for designing apps considering location privacy?
\end{enumerate}

\subsection{Other Privacy-preserving Technologies}
\begin{enumerate}[resume, label=Q\arabic*., wide=0pt, leftmargin=*, font=\itshape]
     \item Have you used any other privacy-preserving technologies? Are you more likely to use privacy-preserving technologies or practices after using the app? Why or why not?
     \item Have you thought about how AI technology/tools help preserve your location privacy?
\end{enumerate}


\end{document}